\documentclass{article}
\usepackage{spconf,amsmath,graphicx,amssymb}
\usepackage{array}

\usepackage{hyperref}
\usepackage{wrapfig}
\usepackage{graphicx}
\usepackage{subfigure}
\usepackage{float}
\usepackage{adjustbox}
\usepackage[justification=centering]{caption}
\usepackage{tabularx}
\usepackage{multirow}
\usepackage[belowskip=-15pt,aboveskip=0pt]{caption}

\newcolumntype{P}[1]{>{\centering\arraybackslash}p{#1}}
\newcolumntype{M}[1]{>{\centering\arraybackslash}m{#1}}
\newcommand{\PreserveBackslash}[1]{\let\temp=\\#1\let\\=\temp}
\newcolumntype{R}[1]{>{\PreserveBackslash\raggedleft}p{#1}}
\usepackage{tabulary,booktabs}
\usepackage{enumitem}
\usepackage{lipsum}
\usepackage{blindtext}
\usepackage[linesnumbered, algo2e, ruled,norelsize]{algorithm2e}
\usepackage{xcolor}
\usepackage{cite}

\SetCommentSty{mycommfont}
\title{Learning-based Lossless Point Cloud Geometry Coding using Sparse Tensors}
\name{Dat Thanh Nguyen and Andr\'e Kaup \thanks{This work was funded by the Deutsche Forschungsgemeinschaft (DFG, German Research Foundation) under Grant SFB 1483 – Project-ID 442419336.}}
\address{Chair of Multimedia Communications and Signal Processing \\ Friedrich-Alexander-Universität Erlangen-Nürnberg (FAU) \\ Erlangen, Germany}

\begin{document}
%
\maketitle
\begin{abstract}
Most point cloud compression methods operate in the voxel or octree domain which is not the original representation of point clouds. Those representations either remove the geometric information or require high computational power for processing. In this paper, we propose a context-based lossless point cloud geometry compression that directly processes the point representation. Operating on a point representation allows us to preserve geometry correlation between points and thus to obtain an accurate context model while significantly reduce the computational cost. Specifically, our method uses a sparse convolution neural network to estimate the voxel occupancy sequentially from the $x,y,z$ input data. Experimental results show that our method outperforms the state-of-the-art geometry compression standard from MPEG with average rate savings of 52\% on a diverse set of point clouds from four different datasets.

\end{abstract}
\section{Introduction}
\label{sec:intro}
\par A Point Cloud (PC) is a set of discrete 3D points that represent 3D scenes or objects. Typical PCs contain millions of points and each point is represented by spatial coordinates ($x,y,z$) and attributes (e.g. color, velocity, etc.) which require a huge amount of storage. Recently, point clouds are becoming the common data structure in many 3D applications, and as a result, efficient point cloud compression (PCC) methods are highly required.

\par The most well-known methodologies were developed by the Moving Picture Expert Group (MPEG) \cite{8571288}. Two approaches have been proposed: Video-based PCC (V-PCC) focuses on dynamic point clouds and Geometry-based PCC (G-PCC) targets static content. V-PCC projects 3D point clouds onto 2D planes and utilizes a 2D video codec to encode the projected videos. On the other hand, G-PCC operates in 3D space and uses 3D tools to encode the point cloud. In most of the methods, geometry and attributes are encoded independently. In this paper, we focus on the point cloud geometry (PCG) compression leaving the compression of attributes for future study.
\par The raw representation of a point cloud geometry is a point representation with millions of $x,y,z$ values. Typically, those coordinate values are first quantized to integer precision prior to geometry coding. At this stage, the point cloud geometry can be converted to a voxel or octree representation. To obtain a voxel representation, the 3D volume is first divided into a fixed number of cubes in each dimension, each cube is called a voxel and a binary value is assigned to each voxel to indicate whether that voxel is occupied or not. The number of voxels per each dimension defines the resolution as well as the bit depth of a voxel block (1024 voxels = 1024 resolution = 10 bits depth). Geometric information (i.e., planes, surfaces, etc.) are preserved in a voxel representation. However, applying conventional operations such as convolution or 3D transforms on a voxel block requires significant computational power as voxel representations bring a large amount of redundancy regardless of the voxels' occupancy state. By recursively splitting a voxel block into eight sub-cubes and using bit 1 to signal an occupied sub-cube, bit 0 to signal an empty sub-cube until the desired precision, we obtain an octree representation. Octree representation lower the geometry bit rate compared to point-based or voxel-based representation, however, geometric information is lost in the octree representation. 
\par In this work, we propose a point-based method for lossless point cloud geometry coding utilizing sparse convolution operation (SparseVoxelDNN). On the one hand, a point-based representation (which we represent using sparse tensors) eliminates the sparsity while retaining the geometric information of the point clouds and thus can be exploited by our context model. On the other hand, with sparse tensor in combination with sparse convolution operation we can easily process point cloud data with low computational complexity. We demonstrate experimentally that our SparseVoxelDNN outperforms G-PCC and the state-of-the-art learning-based geometry compression method by a large margin.


\section{Related work}
\label{sec:stateoftheart}
%
\par Most of the existing point cloud geometry compression methods including MPEG G-PCC are based on an octree structure \cite{schnabel2006octree, garcia2017context, garcia2018intra,huang2020octsqueeze} and a trisoup representation. In the MPEG lossless geometry coder, to encode the octant value of each node, G-PCC exploits correlation within the octree structure by introducing many tools such as neighbour-dependent entropy context \cite{neighbor}, intra prediction \cite{intracodinggpcc}, or planar/angular coding mode \cite{planarcodingmode,angularcodingmode}. MPEG G-PCC is widely used for PCC comparison because of its efficiency and regular updates. 
\par Other context modeling methods based on simpler approaches have been proposed \cite{garcia2017context, garcia2018intra}, aiming at predicting the occupancy pattern using the information from parent nodes, current node position or corresponding nodes at the previous time frame. However, those approaches introduce frequency tables which are collected from the higher octree level or the previous time frame and must be sent to the decoder along with the actual geometry bitstream.

\par Deep learning has been widely applied in image/video compression as well as for point cloud data. Similar to image/video compression, most of the learning-based lossy point cloud geometry compression methods are based on an  auto-encoder architecture  \cite{quach2019learning,wang2019learned,guarda2020point}. Specifically, in \cite{quach2019learning}, 3D voxelized blocks are first encoded into a low dimensional latent space and the reconstruction is cast as a binary classification problem. Recently, an autoregressive-based lossless compression method VoxelDNN has been proposed in \cite{nguyen2021lossless} which was further developed to a multi-scale approach by the authors \cite{nguyen2021multiscale}. Both approaches outperform G-PCC by a large margin on dense point clouds. However, using a voxel representation as input results in high computational costs even with a multi-scale version. 
\par Sparse tensors and sparse convolution have been studied and applied to point cloud tening at reducing the complexity. However, most of the applications are in lossy geometry PCC scenarios \cite{huang20193d, wang2020multiscale,yan2019deep}. In this work, we propose an approach to extract features from a sparse representation and then build an accurate context to estimate the voxel occupancy using autoregressive modeling. Autoregressive models have been proven to be the state-of-the-art likelihood estimation methods \cite{oord2016pixel, salimans2017pixelcnn++}, principally resulting in an accurate occupancy distributions. However, a drawback of autoregressive models is its computational cost. We hence employ sparse tensor and sparse convolution operation to partially reduce the cost.

\section{Proposed method}
\label{proposedmethod}
\par The details of our novel method SparseVoxelDNN will be presented in this section. Our aim is to encode occupancy voxel block $v$ at resolution $d$ containing very few occupied voxels. Instead of processing a voluminous and redundant block $v$, we represent block $v$ in a sparse tensor. We  first  define  a  3D  raster  scan  order  that  scans  voxel  by  voxel  in  depth, height and width order.  For ease of notation, we index all voxels in block $v$ at resolution $d$ from 1 to $d^3$ in raster scan order with:
 \begin{equation}
    v_i= 
    \begin{cases} 
    1, \quad \text{if $i^{th}$ voxel is occupied}\\
    0, \quad \text{otherwise}.
    \end{cases}
\end{equation}
A voxel $v_i$ with $i=x_id^2+y_id+z_i$ is corresponding to a point at $x_i, y_i, z_i$ in sparse tensor. For easier interpretation, we will use $v_i$ to represent a point as well as a voxel in the following sections.
\subsection{Occupancy distribution modeling}
SparseVoxelDNN losslessly encodes a voxel block $v$ of size $d$ using a CABAC coder. Our aims is to build an accurate probability model to predict the joint probability $p(v)$ that can adapt to local occupancy state. We factorize the joint distribution $p(v)$ as a product of conditional distributions over the spatial dimension: 

\begin{equation}
p(v) = \prod_{i=1}^{d^3}p(v_i|v_{i-1},v_{i-2},\ldots,v_{1}). \\
\label{eq:p(v)}
\end{equation}
Each term on the right hand side is the probability of voxel $v_i$ being occupied given the occupancy state of all previous voxels in the 3D raster scan order including $v_{i-1},v_{i-2},\ldots,v_{1}$. In \cite{nguyen2021lossless}, the authors directly estimate $p(v_i|v_{i-1}, \ldots, v_1)$ as binary classification problem using a neural network called VoxelDNN. Instead, inpired by the work from Pixelcnn++ \cite{salimans2017pixelcnn++}, VAE \cite{kingma2016improved}, we model the distribution in a natural way as a continuous distribution, by assuming there is a latent occupancy intensity $\gamma$ with continous distribution, in which the occupancy distribution can be easily extracted. This also enables us to model other channels with a lot of flexibility. Specifically, we model the occupancy intensity of each voxel as a mixture of $L$ logistic distributions which allows us to easily calculate the probability on the observed discretized value as: 

 \begin{equation}
 \begin{split}
    \gamma & \sim \sum_{l=1}^{L}  \pi_l \mathrm{logistic} (\mu_l, s_l) \\
    \hat{p}(v_i=0|\pi,\mu, s)&=\sum_{l=1}^{L}  \pi_l \left [\sigma((x+0.5-\mu_l)/s_l) \right ]\\
    \hat{p}(v_i=1|\pi,\mu, s)&=1-\hat{p}(x=0|\pi,\mu, s)
    \end{split}
    \label{prob}
\end{equation}
where $\pi_l$ is the one hot coding weight which will determine whether the $l^{th}$ distribution to be picked, $\mu_l$ and $s_l$ are the mean and scale parameter of a logistic distribution, while $\sigma()$ is the logistic sigmoid function (i.e. the cdf function of a logistic distribution). The better the distribution estimation is, the smaller will be the coding bitrate we obtain. Therefore, we train our SparseVoxelDNN to estimate $\pi_l$, $\mu_l$ and $s_l$ by minimizing the negative log-likelihood of $\hat{p}$.


\par
\begin{figure}
\captionsetup{singlelinecheck = false, format= hang, justification=justified, font=small, labelsep=space}
\begin{minipage}[b]{.3\linewidth}
  \centering
  \centerline{\includegraphics[width=0.95\linewidth]{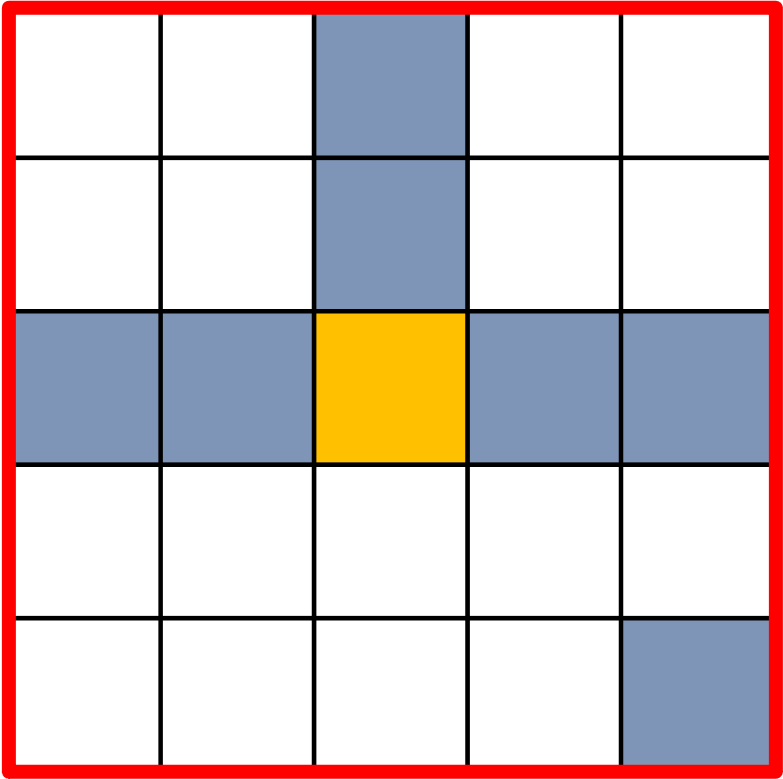}}
  \centerline{(a)}\medskip
\end{minipage}
\hfill
\begin{minipage}[b]{0.3\linewidth}
\label{sfig:typeA}
  \centering
  \centerline{\includegraphics[width=0.95\linewidth]{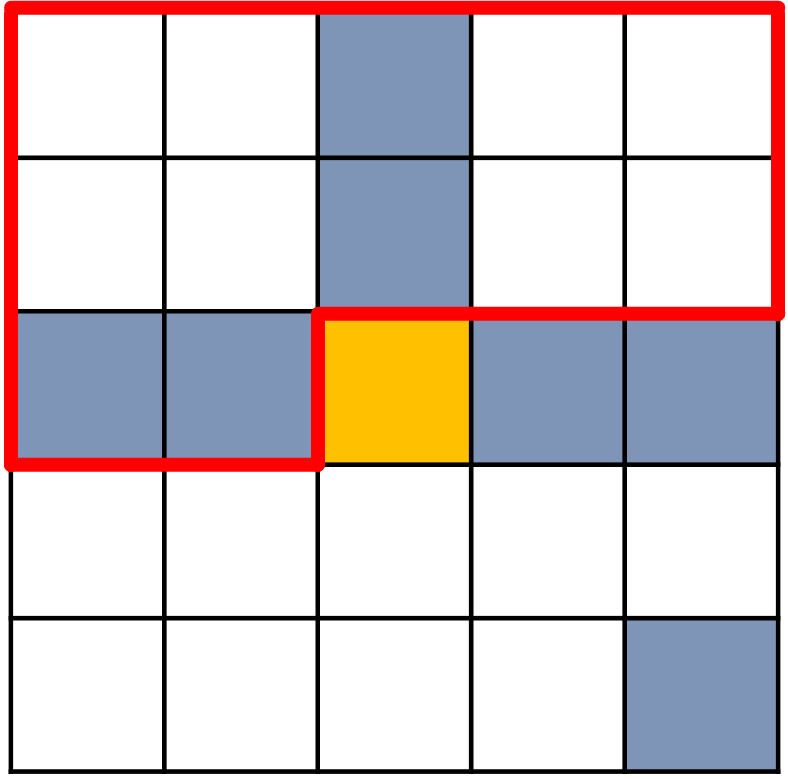}}
  \centerline{(b)}\medskip
\end{minipage}
\hfill
\begin{minipage}[b]{0.3\linewidth}
\label{sfig:typeA}
  \centering
  \centerline{\includegraphics[width=0.95\linewidth]{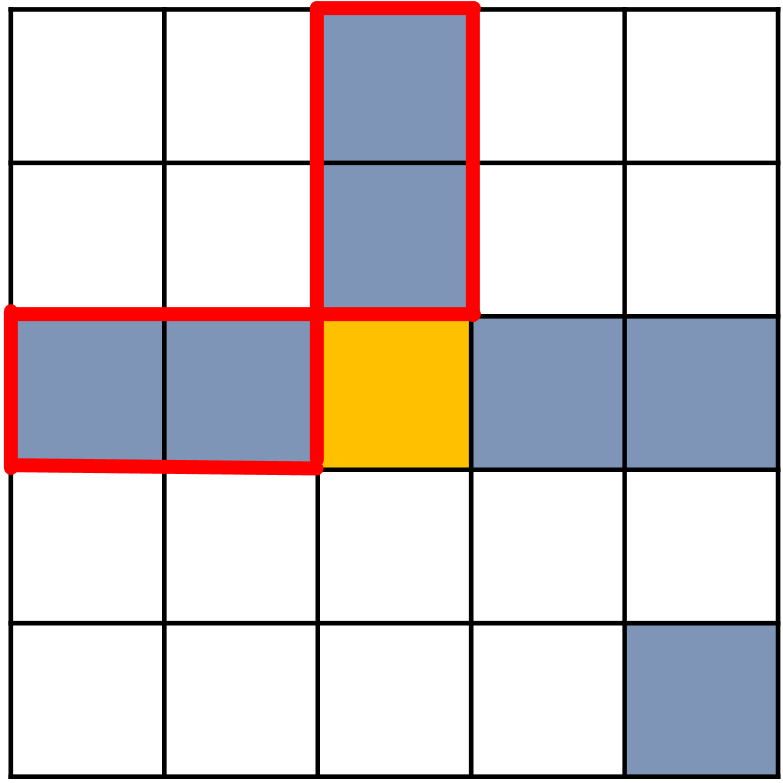}}
  \centerline{(c)}\medskip
\end{minipage}
\caption{Illustration of context in a $5 \times 5$ image. White pixels correspond to empty voxels, grey pixels correspond to occupied voxels and the yellow pixel corresponds to currently processed voxel. The area inside the red boundaries is the receptive field of a convolution operation of size $5\times 5$. (a): Conventional convolution without masks (violating the causality constraint). (b): Conventional convolution with masks \cite{nguyen2021lossless}. (c): Sparse convolution with our sparse masks. }
\label{fig:context}
\end{figure}

 The conditional distribution in (\ref{eq:p(v)}) requires a causality constraint on our SparseVoxelDNN network. In Algorithm \ref{algo:maskalg} we show the type A and type B mask construction which will be applied on sparse convolution filter at the first layer and the subsequent layers to enforce causality, respectively. Note that in sparse convolution implemented by Nvidia \cite{choy20194d} $kernel\_size=k_Dk_Hk_W$  where $k_D, k_H, k_W$ are the 3D kernel sizes.  Simple examples of context in 2D space with and without masks are shown on Figure \ref{fig:context}. Our masks strictly enforce the causality while eliminating all the redundancy within the receptive area and thus providing an accurate context. 

\begin{algorithm2e}[h]
\SetAlgoLined
\SetKwInOut{Input}{Input}
 \small
\KwIn{$mask\_type$} 
$kernel\_size \leftarrow k_D*k_H*k_W$\\
$mask \leftarrow$ $ ones\_like(kernel)$\\
$mask\left[kernel\_size//2+ mask\_type==B,:,: \right]=0$\\
$kernel=kernel*mask$
       
\caption{Mask construction and application}
\label{algo:maskalg}
\end{algorithm2e}

\begin{figure}[tb]
\captionsetup{singlelinecheck = false, format= hang, justification=justified, font=small, labelsep=space}
\centering
\includegraphics[width=0.6\linewidth]{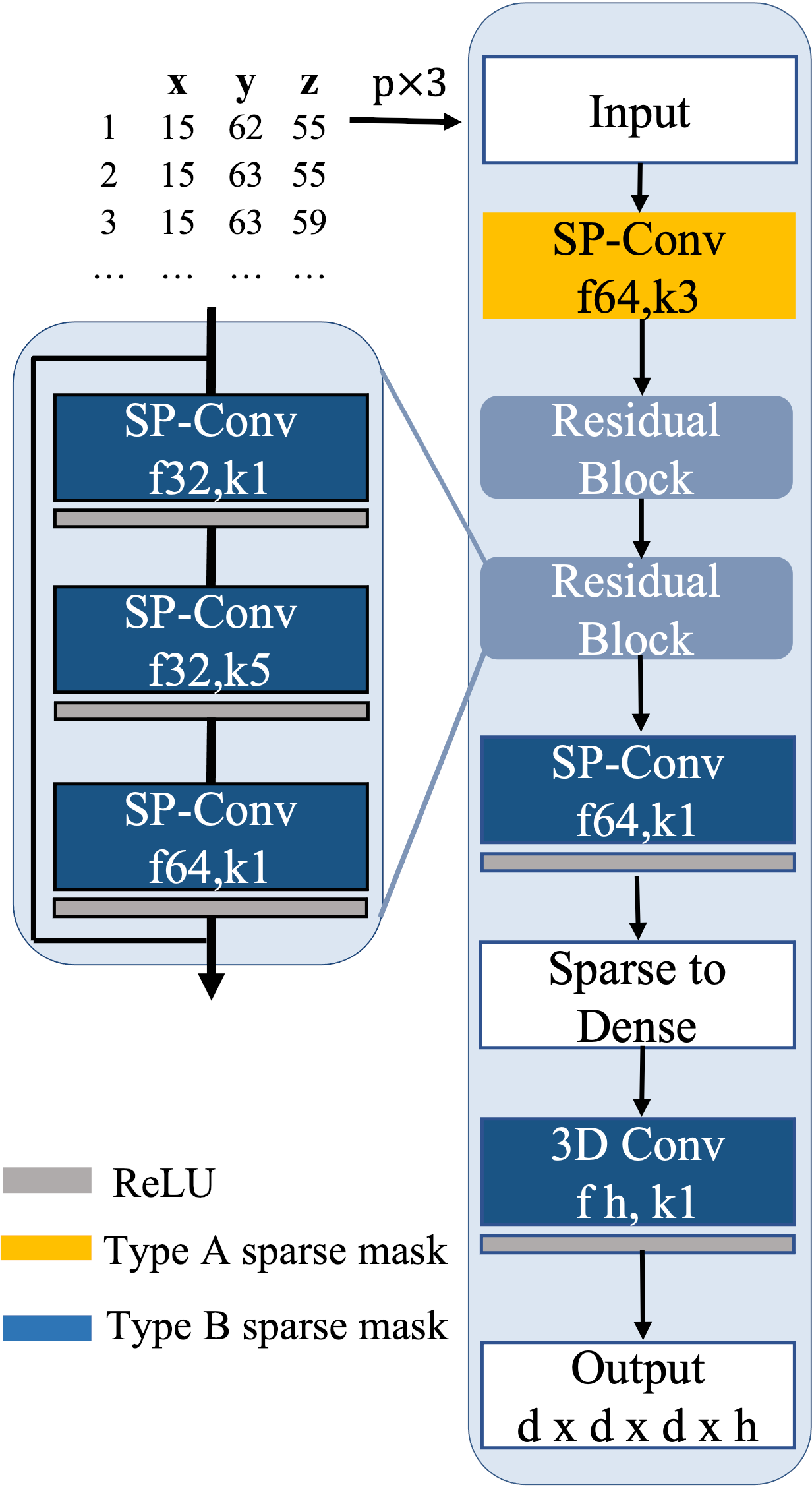}
\caption{SparseVoxelDNN architecture. A sparse type A mask is applied in the first layer (yellow block) and sparse type B masks afterwards. `f64,k3' stands for 64 filters, kernel size 3. White blocks are Input/Output or converting layer and thus do not contain any trainable parameters. }
\label{fig:Networkarchitecture}
\end{figure}

Our SparseVoxelDNN network architecture for a point cloud at resolution $d$ is shown in Figure \ref{fig:Networkarchitecture}. From the first layer to the ``Sparse to Dense'' layer, we use sparse tensor representation. Therefore, we have to make sure that the coordinate corresponding to the next voxel in 3D raster scan order can be generated from the previously decoded coordinates regardless of its occupancy state. We enforce generation of new coordinates on the first layer in such a way that the output coordinates are the outer product of the kernel size and the input coordinates. Next, we employ two residual blocks before converting a point cloud from a sparse tensor to a voxel representation in the ``Sparse to Dense'' layer. The last layer is a $1 \times 1 \times 1$ 3D convolution and thus does not violate the causality. The number of filters $h$ in the last layer is the number of parameter in (\ref{prob}). For example, in our experiments, with a mixture of 5 logistics, we have $h=15$. 

\subsection{SparseVoxelDNN Coder}
\par Ultilizing a sparse tensor does not harm the computational cost as voxel representation but it is still not feasible to process the whole point cloud as a single block. Therefore, prior to the actual encoding process, we partition the point cloud into blocks of size 64 and use an octree to signal the partitioning. By doing this, most of the empty space in the point clouds can be removed. Experimental results show that the octree signal costs less than 2\% of the bitstream. We encode each block of size 64 using our SparseVoxelDNN combined with a CABAC coder. Specifically, we encode voxel by voxel sequentially in 3D raster scan order. Everytime a voxel distribution is obtained, we pass the distribution to an arithmetic coder. The voxel is then fed back to SparseVoxelDNN (in the sparse form) to predict the occupancy distribution of the next voxel. Sequential prediction is a general drawback of all autoregressive-based approaches. We minimize the cost by ultilizing sparse tensor and sparse convolution in our distribution model to have a lighter and faster prediction.
\subsection{Experimental setup}

As training data we consider point clouds from four different datasets, including Microsoft Voxelized Upper Bodies (MVUB) \cite{loop2016microsoft} - dynamic dense voxelized upper-body point clouds, MPEG CAT1 - static sparse point clouds for cultural heritage and other 3D photography applications, MPEG Owlii \cite{owlii} and 8i \cite{d20178i} - dynamic dense voxelized full-body point clouds. To enforce the fairness between the  datasets in which we select point clouds for testing, point clouds from MPEG CAT1 are sampled to 10 bit precision as in MVUB, Owlii and 8i. To train a SparseVoxelDNN model of size $d=64$, we divide all selected PCs into occupied blocks of size $d \times d \times d$. Table \ref{table:noblocks} reports the number of blocks from each dataset for training.
\begin{table}[t]
\caption{Number of blocks in the training sets of each model}
\centering
\resizebox{0.9\linewidth}{!}{
\begin{tabular}{lR{0.8cm}R{0.8cm}R{0.8cm}R{1.2cm}R{1cm}}
\hline
\begin{bf}  \end{bf}
&\begin{bf}MVUB\end{bf}
& \begin{bf}8i\end{bf} 
&\begin{bf}CAT1\end{bf}
& \begin{bf}Owlii\end{bf} 
& \begin{bf}Total\end{bf} \\
\hline
Block 64 & 317763 &235369 & 2499 & 358834 & 914465\\
\hline
\vspace{0.1mm}
\end{tabular}}
\label{table:noblocks}
\end{table}
\par We add data augmentation techniques in the training phase to increase the generalization of our network. Specifically, rotation and sub-sampling are applied on training blocks. The training and testing process are performed on a a GeForce RTX 3090 GPU with Adam optimizer with learning rate $10^{-4}$ and accumulating gradients. We set the batch size to 8, accumulating step to 16 and early stopping callback with patience of 5 epochs. We evaluate our geometry compression methods on a set of 10 bit point clouds from MVUB, 8i, CAT1 which are not yet used in the training phase and two inanimate sparse point clouds from the University of S\~ao Paulo, Brazil (USP) \cite{usp}.

\section{Experimental Results}
\label{performanceeval}

\begin{table}[t]
\caption{Comparions of our method against G-PCC v14 and VoxelDNN}
\resizebox{0.99\linewidth}{!}{ \begin{tabular}{|P{0.1cm}|l|P{1.2cm}||R{0.7cm}|R{1.4cm}|R{0.7cm}|R{1.4cm}|}
\cline{3-7}
\multicolumn{2}{c|}{}
& \begin{bf} G-PCC \end{bf}
& \multicolumn{2}{c|}{\begin{bf}VoxelDNN\end{bf}}
& \multicolumn{2}{c|}{\begin{bf}SparseVoxelDNN\end{bf}}
\\
\cline{2-7}
\multicolumn{1}{c|}{}&Point Cloud & bpov&bpov&Gain over G-PCC &bpov&Gain over G-PCC\\
\hline

\multirow{3}{*}{\rotatebox[origin=c]{90}{MVUB}}&Phil& 1.15 &0.82 &-28\%& 0.40 &-66\%  \\ 
\cline{2-7}
&Ricardo&1.07 &0.75 &-29\%& 0.35&-67\% \\
\cline{2-7}
&\textbf{Average}  &\textbf{1.11}  &\textbf{0.79} &\textbf{-28.5\%} & \textbf{0.37}&\textbf{-66.4\%}  \\
\cline{2-7}
\hline

\multirow{5}{*}{\rotatebox[origin=c]{90}{8i}}&Redandblack &1.09  &0.70 &-36\%& 0.30&-64\% \\
\cline{2-7}
&Loot &0.95&0.61 &-36\% &0.33 & -65\%\\
\cline{2-7}
&Thaidancer&1.00 &0.66 &-34\%&0.32 &-68\% \\
\cline{2-7}
&Boxer&0.95  &0.59 &-38\%& 0.30&-68\% \\
\cline{2-7}
&\textbf{Average} &\textbf{1.00} &\textbf{0.64}& \textbf{-36.0\%}&\textbf{0.34} &\textbf{-66.5\%} \\
\hline

\multirow{4}{*}{\rotatebox[origin=c]{90}{CAT1}}&Frog &1.92   &1.70 &-10\%& 1.23&-36\%  \\
\cline{2-7}
&Arco Valentino &4.85 &4.99 &+3\%& 3.03&-38\% \\
\cline{2-7}
&Shiva&3.67 & 3.50 &-4\%&2.66&-28\% \\
\cline{2-7}
&\textbf{Average} &\textbf{3.48} &\textbf{3.40} &\textbf{-3.8\%}&\textbf{2.31} &\textbf{-33.6\%}\\
\hline

\multirow{3}{*}{\rotatebox[origin=c]{90}{USP}}&BumbaMeuBoi&5.41   &5.07 &-6\%& 2.52&-53\%  \\
\cline{2-7}
&RomanOiLight& 1.86  &1.62 &-13\%& 1.52&-19\% \\
\cline{2-7}
&\textbf{Average}  &\textbf{3.64}  &\textbf{3.35} &\textbf{-9.6\%}&\textbf{2.02}&\textbf{-35.8\%}  \\
\cline{2-7}

\hline
\hline
\end{tabular}}
\label{table:resulttable}
\vskip 3mm  
\end{table}

\par We compare the encoding bitrate measured in average bit per occupied voxel (bpov) of our method with MPEG G-PCC v14\cite{8571288} and the state-of-the-art autoregressive-based method VoxelDNN \cite{nguyen2021lossless}. Table \ref{table:resulttable} reports the $bpov$ of these approaches as well as the gain of SparseVoxelDNN and VoxelDNN over G-PCC for each point cloud and the average per dataset.

\par First, we observe that SparseVoxelDNN outperforms both G-PCC and VoxelDNN by a large margin in all test point clouds. Our method works better on dense point clouds than sparse point clouds with around $66\%$ of bitrate reduction on MVUB and 8i. This could be explained by the fact that the context model can efficiently exploit the relation between points and predict a more accurate probability on dense point clouds than on sparse point clouds. However, despite the different content type and density of input point clouds, we obtain a significant bitrate reduction of $33.6\%$ and $35.8\%$ in average on CAT1 and USP compared to G-PCC, respectively. Compared to VoxelDNN, our method obtains, on average, a further bitrate reduction of $29\%$ on all tested point clouds. VoxelDNN does not achieve a clear gain on sparse point clouds from CAT1 and USP, as voxel blocks contain a large amount of redundancy and thus the context model is less efficiency compared to the context model with sparse representation (as shown in Figure \ref{fig:context}).

\par Table \ref{table:complexity} reports the encoding, decoding time in seconds of our method, G-PCC, and VoxelDNN \cite{nguyen2021lossless}. Our approach obtains a significant bitrate gain over G-PCC and the price to be paid for this performance improvement is a higher computational cost with average encoding time of  $4.7 \times$ over G-PCC. Compared to VoxelDNN, SparseVoxelDNN is $49 \times$  faster in encoding runtime. There is still the common decoding complexity issue of autoregressive-based approaches (VoxelDNN and SparseVoxelDNN), however the problem can be alleviated with multiscale structures as disscussed in \cite{nguyen2021multiscale, salimans2017pixelcnn++}.

\begin{table}[t]
\caption{Average runtime (in seconds) of different encoders comparing with G-PCC}
\centering
\resizebox{0.99\linewidth}{!}{ \begin{tabular}{M{0.7cm}M{1.8cm}M{3.5cm}M{2.5cm}}
\hline
\begin{bf}  \end{bf}
&\begin{bf}G-PCC\end{bf}
& \begin{bf}VoxelDNN\end{bf} 
& \begin{bf}SparseVoxelDNN\end{bf} \\
\hline
(Enc) &1.6&  355&7.2\\
\hline
(Dec) &1.1&  330&229\\
\hline
\end{tabular}}
\label{table:complexity}
\end{table}
\section{Conclusions}
\label{conclusion}
\par In this paper, we have presented a novel lossless point cloud geometry compression method called SparseVoxelDNN. We represent the point cloud using both octree, voxel and point representation in a way that is most suitable to their characteristics. We build a 3D autoregressive model to predict continuous occupancy distribution by deploying 3D sparse convolution masks in a point-based neural network. This approach enables us to accurately model the occupancy distribution and obtain significant gains, with average rate savings of $52\%$ over G-PCC while avoiding expensive computations. We are now developing lossy point cloud compression methods using sparse convolution and sparse tensors.




\bibliographystyle{IEEEbib}
\bibliography{references/IEEEabrv.bib, references/refs.bib}

\end{document}